\documentstyle[12pt]{article}

\catcode`\@=11
\long\def\@makefntext#1{ 
\protect\noindent \hbox to 3.2pt {\hskip-.9pt
$^{{\ninerm\@thefnmark}}$\hfil}#1\hfill} 

\def\thefootnote{\fnsymbol{footnote}}
 \def\@makefnmark{\hbox to 0pt{$^{\@thefnmark}$\hss}}  

\def\ps@myheadings{\let\@mkboth\@gobbletwo
\def\@oddhead{\hbox{} 
\rightmark\hfil\ninerm\thepage}
\def\@oddfoot{}\def\@evenhead{\ninerm\thepage\hfil 
\leftmark\hbox{}}\def\@evenfoot{}
\def\sectionmark##1{}\def\subsectionmark##1{}}

\begin{document}

\input epsf

\newcommand{\symbolfootnote}{\renewcommand{\thefootnote}
	{\fnsymbol{footnote}}}
\renewcommand{\thefootnote}{\fnsymbol{footnote}}
\newcommand{\alphfootnote}
	{\setcounter{footnote}{0}
	 \renewcommand{\thefootnote}{\sevenrm\alph{footnote}}}

\newcounter{sectionc}\newcounter{subsectionc}\newcounter{subsubsectionc}
\renewcommand{\section}[1] {\vspace{0.6cm}\addtocounter{sectionc}{1}
\setcounter{subsectionc}{0}\setcounter{subsubsectionc}{0}\noindent
	{\bf\thesectionc. #1}\par\vspace{0.4cm}}
\renewcommand{\subsection}[1] {\vspace{0.6cm}\addtocounter{subsectionc}{1}
	\setcounter{subsubsectionc}{0}\noindent
	{\it\thesectionc.\thesubsectionc. #1}\par\vspace{0.4cm}}
\renewcommand{\subsubsection}[1]
{\vspace{0.6cm}\addtocounter{subsubsectionc}{1}
	\noindent {\rm\thesectionc.\thesubsectionc.\thesubsubsectionc.
	#1}\par\vspace{0.4cm}}
\newcommand{\nonumsection}[1] {\vspace{0.6cm}\noindent{\bf #1}
	\par\vspace{0.4cm}}

\newcounter{appendixc}
\newcounter{subappendixc}[appendixc]
\newcounter{subsubappendixc}[subappendixc]
\renewcommand{\thesubappendixc}{\Alph{appendixc}.\arabic{subappendixc}}
\renewcommand{\thesubsubappendixc}
	{\Alph{appendixc}.\arabic{subappendixc}.\arabic{subsubappendixc}}

\renewcommand{\appendix}[1] {\vspace{0.6cm}
	\refstepcounter{appendixc}
	\setcounter{figure}{0}
	\setcounter{table}{0}
	\setcounter{equation}{0}
	\renewcommand{\thefigure}{\Alph{appendixc}.\arabic{figure}}
	\renewcommand{\thetable}{\Alph{appendixc}.\arabic{table}}
	\renewcommand{\theappendixc}{\Alph{appendixc}}
	\renewcommand{\theequation}{\Alph{appendixc}.\arabic{equation}}
	\noindent{\bf Appendix \theappendixc #1}\par\vspace{0.4cm}}
\newcommand{\subappendix}[1] {\vspace{0.6cm}
	\refstepcounter{subappendixc}
	\noindent{\bf Appendix \thesubappendixc. #1}\par\vspace{0.4cm}}
\newcommand{\subsubappendix}[1] {\vspace{0.6cm}
	\refstepcounter{subsubappendixc}
	\noindent{\it Appendix \thesubsubappendixc. #1}
	\par\vspace{0.4cm}}

\def\abstracts#1{{
	\centering{\begin{minipage}{30pc}\tenrm\baselineskip=12pt\noindent
	\centerline{\tenrm ABSTRACT}\vspace{0.3cm}
	\parindent=0pt #1
	\end{minipage} }\par}}

\newcommand{\bibit}{\it}
\newcommand{\bibbf}{\bf}
\renewenvironment{thebibliography}[1]
	{\begin{list}{\arabic{enumi}.}
	{\usecounter{enumi}\setlength{\parsep}{0pt}
\setlength{\leftmargin 1.25cm}{\rightmargin 0pt}
	 \setlength{\itemsep}{0pt} \settowidth
	{\labelwidth}{#1.}\sloppy}}{\end{list}}

\topsep=0in\parsep=0in\itemsep=0in
\parindent=1.5pc

\newcounter{itemlistc}
\newcounter{romanlistc}
\newcounter{alphlistc}
\newcounter{arabiclistc}
\newenvironment{itemlist}
	{\setcounter{itemlistc}{0}
	 \begin{list}{$\bullet$}
	{\usecounter{itemlistc}
	 \setlength{\parsep}{0pt}
	 \setlength{\itemsep}{0pt}}}{\end{list}}

\newenvironment{romanlist}
	{\setcounter{romanlistc}{0}
	 \begin{list}{$($\roman{romanlistc}$)$}
	{\usecounter{romanlistc}
	 \setlength{\parsep}{0pt}
	 \setlength{\itemsep}{0pt}}}{\end{list}}

\newenvironment{alphlist}
	{\setcounter{alphlistc}{0}
	 \begin{list}{$($\alph{alphlistc}$)$}
	{\usecounter{alphlistc}
	 \setlength{\parsep}{0pt}
	 \setlength{\itemsep}{0pt}}}{\end{list}}

\newenvironment{arabiclist}
	{\setcounter{arabiclistc}{0}
	 \begin{list}{\arabic{arabiclistc}}
	{\usecounter{arabiclistc}
	 \setlength{\parsep}{0pt}
	 \setlength{\itemsep}{0pt}}}{\end{list}}

\newcommand{\fcaption}[1]{
	\refstepcounter{figure}
	\setbox\@tempboxa = \hbox{\tenrm Fig.~\thefigure. #1}
	\ifdim \wd\@tempboxa > 6in
	   {\begin{center}
	\parbox{6in}{\tenrm\baselineskip=12pt Fig.~\thefigure. #1 }
	    \end{center}}
	\else
	     {\begin{center}
	     {\tenrm Fig.~\thefigure. #1}
	      \end{center}}
	\fi}

\newcommand{\tcaption}[1]{
	\refstepcounter{table}
	\setbox\@tempboxa = \hbox{\tenrm Table~\thetable. #1}
	\ifdim \wd\@tempboxa > 6in
	   {\begin{center}
	\parbox{6in}{\tenrm\baselineskip=12pt Table~\thetable. #1 }
	    \end{center}}
	\else
	     {\begin{center}
	     {\tenrm Table~\thetable. #1}
	      \end{center}}
	\fi}

\def\@citex[#1]#2{\if@filesw\immediate\write\@auxout
	{\string\citation{#2}}\fi
\def\@citea{}\@cite{\@for\@citeb:=#2\do
	{\@citea\def\@citea{,}\@ifundefined
	{b@\@citeb}{{\bf ?}\@warning
	{Citation `\@citeb' on page \thepage \space undefined}}
	{\csname b@\@citeb\endcsname}}}{#1}}

\newif\if@cghi
\def\cite{\@cghitrue\@ifnextchar [{\@tempswatrue
	\@citex}{\@tempswafalse\@citex[]}}
\def\citelow{\@cghifalse\@ifnextchar [{\@tempswatrue
	\@citex}{\@tempswafalse\@citex[]}}
\def\@cite#1#2{{$\null^{#1}$\if@tempswa\typeout
	{IJCGA warning: optional citation argument
	ignored: `#2'} \fi}}
\newcommand{\citeup}{\cite}

\newcommand{\be}{\begin{equation}}
\newcommand{\ee}{\end{equation}}
\newcommand{\bea}{\begin{eqnarray}}
\newcommand{\eea}{\end{eqnarray}}
\newcommand{\sumint}{\hbox{$\sum$}\!\!\!\!\!\!\!\int}
\newcommand{\cm}{M}
\def\fnm#1{$^{\mbox{\scriptsize #1}}$}
\def\fnt#1#2{\footnotetext{\kern-.3em
	{$^{\mbox{\sevenrm #1}}$}{#2}}}

\font\twelvebf=cmbx10 scaled\magstep 1
\font\twelverm=cmr10 scaled\magstep 1
\font\twelveit=cmti10 scaled\magstep 1
\font\elevenbfit=cmbxti10 scaled\magstephalf
\font\elevenbf=cmbx10 scaled\magstephalf
\font\elevenrm=cmr10 scaled\magstephalf
\font\elevenit=cmti10 scaled\magstephalf
\font\bfit=cmbxti10
\font\tenbf=cmbx10
\font\tenrm=cmr10
\font\tenit=cmti10
\font\ninebf=cmbx9
\font\ninerm=cmr9
\font\nineit=cmti9
\font\eightbf=cmbx8
\font\eightrm=cmr8
\font\eightit=cmti8


\centerline{\tenbf ON THE NATURE OF THE ABELIAN HIGGS MODEL
PHASE TRANSITION}
\vspace{0.8cm}
\centerline{\tenrm Giovanni AMELINO-CAMELIA}
\baselineskip=13pt
\centerline{\tenit Center for Theoretical Physics, Laboratory
for Nuclear Science,}
\baselineskip=12pt
\centerline{\tenit and Department of Physics,
Massachusetts Institute of Technology}
\baselineskip=12pt
\centerline{\tenit Cambridge, Massachusetts 02139, USA}
\vspace{0.9cm}
\abstracts{The nature of the Abelian Higgs Model phase transition
is investigated.
A variational approximation is used in the evaluation of the
relevant finite temperature effective potential.
Some of the results presented
are valid not only in the Abelian Higgs Model, but
also in more complex theories.
}

\vfil
\vspace{1.cm}
\twelverm   
\baselineskip=14pt

Several recent studies$^{1-6}$ have been devoted to the phase transition
of the Abelian Higgs Model. Besides being interesting in its own right,
the Abelian Higgs Model gives a simple setting
in which one can develop techniques that might be
useful in the investigation of other gauge theories
with spontaneous symmetry breaking.
In this lecture I discuss
a variational technique that can be used in the evaluation
of the finite temperature effective potential,
which is an important tool in the investigation of phase transitions.
I illustrate this technique by studying the
daisy and superdaisy resummed\cite{gacpi,pap8,hsu}
finite temperature effective potential
of the Abelian Higgs Model.

As discussed in Ref.8, using the composite operator
method\cite{gacpi,pap8,corn} one can show that, if $e^3 < \lambda < e^2$,
the daisy and superdaisy resummed effective potential
of the Abelian Higgs Model can be written as
\bea
V^{res}_T \!\!\! &=& \!\!\! V^{res}_T(\phi,G_0) =
-{m^2 \over 2} \phi^2 + {\lambda \over 24} \phi^4
+{1 \over 2} \, \sumint_k ~ \ln G_0^{-1}(k)
+ {1 \over 2} \, \sumint_k ~ [D^{-1}(\phi;k) G_0(k) -1]
\nonumber\\
& & \!\! + V_2^{(a)}(G_0)
+ V_2^{(b)}(G_0)
+ V_2^{(c)}(G_0)
+ V_2^{(d)}(G_0)
{}~,\label{hfahm}
\eea

\noindent
where
\be
V_2^{(a)}(G) \equiv - {\lambda \over 4!} \sumint_p  \sumint_q \,\,
[G_{aa}(p) G_{bb}(q) + 2 G_{ab}(p) G_{ba}(q)]
{}~,\label{hfahmb}
\ee
\be
V_2^{(b)}(G) \equiv - {e^2 \over 2} g^{\mu \nu}
\sumint_p  \sumint_q \,
G_{\mu \nu}(q) G_{aa}(p)
{}~,\label{hfahmc}
\ee
\be
V_2^{(c)}(G) \equiv {e^2 \over 4} \sumint_p  \sumint_q \,
\epsilon_{ab} \epsilon_{cd} (2p+q)^\mu (2p+q)^\nu
G_{\mu \nu}(q) G_{ad}(p) G_{bc}(p+q)
{}~,\label{hfahmd}
\ee
\be
V_2^{(d)}(G) \equiv {e^2 \over 2} \sumint_p  \sumint_q \,
 \epsilon_{ac} \epsilon_{db} (2q+p)^\mu (2q+p)^\nu
G_{ab}(p+q) G_{c \nu}(p) G_{\mu d}(q)
{}~.\label{hfahme}
\ee

\noindent
$D$ is the tree-level propagator in momentum space, which
can be written as
\bea
(D^{-1}(\phi; k))_{\mu \nu}&=&
(e^2 \phi^2 - k^2)
({k_{\mu} k_{\nu} \over k^2}
- g_{\mu \nu}  ) +
({k^2 \over \xi} - e^2 \phi^2) {k_{\mu} k_{\nu} \over k^2} \nonumber\\
(D^{-1}(\phi ;k))_{ab}&=&
({\lambda \phi^2 \over 2} - m^2) \delta_{a1} \delta_{b1} \, + \,
({\lambda \phi^2 \over 6} - m^2) \delta_{a2} \delta_{b2}
- \delta_{ab} k^2 \nonumber\\
(D^{-1}(\phi ;k))_{a \mu}&=& - i e k_{\mu} \epsilon_{ab} \phi_b
{}~,\label{protre}
\eea

\noindent
and $G_0$ is the solution of
\be
{\delta V^{res}_T(\phi,G) \over \delta G}=0 ~.\label{statahm}
\ee

The analytic solution of Eq.(\ref{statahm}) is beyond our present
technical capabilities, and, as a consequence,
we cannot evaluate $V_T^{res}$ exactly, unless we resort
to numerical methods.

I study $V_T^{res}$ analytically using
the observation that
an approximate solution of the variational
problem (\ref{hfahm})-(\ref{statahm}) can be obtained by
evaluating $V_T^{res}(\phi,G)$ with
specific parameter-dependent expressions for
$G(k)$ and then varying these parameters.
This type of procedure is known\cite{corn}
as the ``Rayleigh-Ritz variational approximation''.
As the parameter-dependent $G(k)$ I take the following
expressions
\bea
G^{-1}_{\mu \nu} &=&
(M^2_t - k^2) t_{\mu \nu}(k) +
(M^2_l - k^2) l_{\mu \nu}(k) +
({k^2 \over \xi} - e^2 \phi^2) {k_{\mu} k_{\nu} \over k^2}
{}~,\nonumber\\
G^{-1}_{ab} &=& \delta_{a1} \delta_{b1} (M_\phi^2 - k^2)+
\delta_{a2} \delta_{b2} (M_\chi^2 - k^2)
{}~,\nonumber\\
G^{-1}_{a \mu} &=& - i e k_{\mu} \epsilon_{ab} \phi_b
{}~.\label{ansrr}
\eea

\noindent
where $t_{\mu \nu}$ and $l_{\mu \nu}$ are defined by
\be
t_{\mu \nu}(k) \equiv
\delta_{\mu i} \delta_{\nu j}  (\delta^{ij} - {k^i k^j \over {\bf k}^2})
{}~,~~
l_{\mu \nu}(k) \equiv
{k_\mu k_\nu \over k^2} - g_{\mu \nu}  - t_{\mu \nu}
{}~.\label{ptpl}
\ee

\noindent
The Eqs.(\ref{ansrr})
express the propagator in terms of
``Rayleigh-Ritz effective masses" $M_x$.
[As required by the way Lorentz invariance is
broken at finite temperature\cite{pisarsk},
the two transverse modes of the
gauge boson acquire the same effective
mass $M_t$ whereas the longitudinal mode has an independent
effective mass $M_l$.]

The approximation
of the daisy and superdaisy resummed finite temperature effective potential
that I evaluate is the solution of the following variational problem
\be
V^{res}_T \simeq V^{res}_T(\phi,G(\{M_0\}))
{}~,\label{rra}
\ee
\be
\biggl[{\delta V^{res}_T(\phi,G(\{M\} ))
\over \delta M^n}\biggr]_{\{ M \}=\{M_0 \}}
= 0
{}~,\label{rrb}
\ee

\noindent
where $\{M \} \!\! \equiv \!\! \{M^1,M^2,M^3,M^4 \} \!\! \equiv \!\!
\{ M_\phi, M_\chi, M_t, M_l \}$.

The effective potential $V^{res}_T(\phi,G(\{M\} ))$
in Eqs.(\ref{rra})-(\ref{rrb}) includes divergent integrals; therefore a
regularization and renormalization procedure is necessary.
In the similar renormalization
of the $\lambda \Phi^4$ scalar theory\cite{gacpi} it has been shown that
the only effect of renormalization
on the high-temperature part of the effective potential is the substitution
of the bare parameters with renormalized ones.
In the following I shall assume that the same applies in the case of the
Abelian Higgs Model,
and therefore, rather than performing renormalization
explicitly, I shall simply omit the (zero-temperature)
ultraviolet-divergent contributions and
substitute the
bare parameters with renormalized ones in my high-temperature
effective potential.

Using the well-known results\cite{doja}
\bea
 \sumint_k \, \ln[k^2-y^2]
\!\!&\simeq& \!\!  -{ \pi^2 T^4 \over 45} + {y^2 T^2 \over 12}
- {y^3 T \over 6 \pi}
+ {c_\Omega y^4 \over 16 \pi^2} ~, \nonumber\\
\sumint_k \, {1 \over k^2 - y^2 }
\!\!&\simeq& \!\! {T^2 \over 12} - {T y \over 4 \pi}
+ {c_\Omega \over 8 \pi^2} y^2 ~,
\nonumber\\
c_\Omega \!\!&\equiv& \!\!  {1 \over 2} \ln({T^2 \over \mu^2})
+{1 \over 2}+ \ln(4 \pi) - \gamma_{Euler} ~,
\label{tadp}
\eea

\noindent
(where $\mu$ is a renormalization scale),
and the high-temperature
approximation of $V_2^{(c)}$ obtained in Ref.8, one can easily show
that, for $M_x^2/T^2 \! << \! 1$, $V^{res}_T(\phi,G(\{M\} ))$ can be
approximated by
\bea
V^{res}_T(\phi,G(\{ M \})) & \simeq &
-{1 \over 2} m^2 \phi^2 + {\lambda \over 24} \phi^4
+{T^2  \over 24}
({3 \over 2} \lambda \phi^2 - 2 m^2 + 3 e^2 \phi^2) \nonumber\\
& & + {T  \over 24 \pi} (M^3_{\phi}+M^3_{\chi}+ 2 M^3_{t}+M^3_{l})
\nonumber\\
& & -{c_\Omega  \over 32 \pi^2}
(M^4_{\phi}+M^4_{\chi}+ 2 M^4_{t}+M^4_{l}) \nonumber\\
& & + {e^2 T^2 \over  32 \pi^2} (M_t^2 -2 M^2_\phi - 2 M^2_\chi)
\ln({M_t + M_\phi + M_\chi \over 3 T})
\nonumber\\
& & - {e^2 T \phi^2 \over 8 \pi} (2 M_t + M_l)
- {T \over 8 \pi} [ M_\phi ({\lambda _\phi^2 \over 2} - m^2)
+ M_\chi ({\lambda _\phi^2 \over 6} - m^2) ]
\nonumber\\
& & - {e^2 T^3 \over  24 \pi} M_l
- ({\lambda \over 144 \pi} + {e^2 \over  32 \pi^2})
T^3 (M_\phi + M_\chi)
\nonumber\\
& & + e^2 T^2 [ {a_{\Phi}  \over  16 \pi} (M^2_{\phi}+M^2_{\chi})
- {c_{\Theta t} \over 128 \pi^2} M_t^2
+ ({c_\Omega \over 48 \pi^2} - {c_{\Theta l} \over 128 \pi^2}) M_l^2 ]
\nonumber\\
& & + {e^2 T^2  \over 32 \pi^2} (M_{l} + M_t) (M_{\phi}+M_{\chi})
+  ({e^2 \over 32} + {\lambda \over 192})  { T^2  \over  \pi^2}
M_{\phi} M_{\chi}
{}~,\label{vresint}
\eea

\noindent
where $a_\Phi \!\! \equiv \!\!
[(c_\Omega + c_{\Theta t} + c_{\Theta l} )/(4 \pi)]
+\lambda  [(4 c_\Omega + 9)/(72 \pi e^2)]$.
$c_{\Theta t}$ and $c_{\Theta l}$,
which are coefficients analogous to $c_\Omega$,
are given by integrals
that can be evaluated numerically\cite{pap8,parwani}.

The approximation of $V^{res}_T(\phi,G(\{ M \}))$
obtained in Eq.(\ref{vresint})
allows to express the gap equations (\ref{rrb})
in the following high-temperature form
\bea
M^2_{\phi(\chi)} & \simeq & m^2_{\phi(\chi)}
+({ \lambda \over 18} + {e^2 \over 4}) T^2
-\biggl[a-{1 \over 4 \pi}
-{1 \over \pi} \ln ({M_\phi + M_\chi \over 3 T}) \biggr]
e^2 T M_{\phi(\chi)} \nonumber\\
& & - { \lambda \over 24 \pi} T M_{\chi(\phi)}
- {e^2 \over 4 \pi} T M_l
+{ c_\Omega  \over  \pi} { M^3_{\phi(\chi)}  \over T}
{}~,\label{appgapma}
\eea
\be
M^2_t \simeq  e^2 \phi^2
+\biggl[ {c_{\Theta t} \over 16 \pi} -{1 \over 8 \pi}
-{1 \over 4 \pi} \ln ({M_\phi + M_\chi \over 3 T}) \biggr]
e^2 T M_t +{ c_\Omega  \over  \pi} { M^3_t  \over T}
{}~,\label{appgapmb}
\ee
\be
M^2_l \simeq e^2 \phi^2 + {e^2 \over 3} T^2
- e^2 ({c_\Omega \over 3 \pi} - {c_{\Theta l} \over 8 \pi}) T M_{l}
- {e^2 T \over 4 \pi} (M_{\phi} + M_{\chi})
+{ c_\Omega  \over  \pi} { M^3_l  \over T}
{}~.\label{appgapmc}
\ee

Finally, reexpressing some terms in Eq.(\ref{vresint}) using the gap
equations (\ref{appgapma})-(\ref{appgapmc}), I find that
the Rayleigh-Ritz
and high-temperature approximation of the daisy and superdaisy resummed
finite temperature effective potential for the
Abelian Higgs Model is given by
\bea
V^{res}_T(\phi,\{ M_0 \}) & \simeq &
-{1 \over 2} m^2 \phi^2 + {\lambda \over 4!} \phi^4
+{ T^2  \over 24}
({2 \over 3} \lambda \phi^2 - 2 m^2 + 3 e^2 \phi^2) \nonumber\\
& & - { T  \over 12 \pi} (M^3_{\phi,0}+M^3_{\chi,0}+ 2 M^3_{t,0}+M^3_{l,0})
\nonumber\\
& & + {e^2 T^2 (2 M^2_{\phi,0} + 2 M^2_{\chi,0} - M_{t,0}^2) \over  32 \pi^2}
\ln({M_{\phi,0} + M_{\chi,0} \over 3 T}) \nonumber\\
& & + {3 c_\Omega  \over 32 \pi^2}
(M^4_{\phi,0}+M^4_{\chi,0}+ 2 M^4_{t,0}+M^4_{l,0}) \nonumber\\
& & - {e^2 T^2  \over 32 \pi^2} M_{l,0} (M_{\phi,0}+M_{\chi,0})
+  ({e^2 \over 32} - {\lambda \over 192})  { T^2  \over  \pi^2}
M_{\phi,0} M_{\chi,0} \nonumber\\
& & - {e^2 T^2  \over  \pi^2}
({c_{\Omega}  \over 48 } + {c_{\Theta t}  \over 128}) M_{l,0}^2
+\tilde{a}_\Phi {e^2 T^2  \over  \pi^2} (M^2_{\phi,0}+M^2_{\chi,0})
\nonumber\\
& & + {c_{\Theta t}  \over 128 \pi^2} e^2 T^2 M_{t,0}^2
{}~,\label{vresult}
\eea

\noindent
where $M_{\phi,0}$, $M_{\chi,0}$, $M_{t,0}$, and $M_{l,0}$ are the solutions
of the gap equations (\ref{appgapma})-(\ref{appgapmc}), and
$\tilde{a}_\Phi \!\! \equiv \!\! 1/32 - (c_{\Theta t}+c_{\Theta l})/64
+(\lambda / e^2)(c_\Omega /288 - 1/128)$.

Concerning the nature of the phase transition of the Abelian Higgs Model
it is useful to notice that for $e T << \phi << T$
the Eqs.(\ref{appgapma})-(\ref{vresult}) imply
that: (I) besides the expected contributions involving even
powers of $\phi$, there
is a negative contribution of order $e^3 T \phi^3$
to the effective potential, which
comes from the $T M_{t,0}^3$ term,
and (II) there is no contribution of order $e^3 T^3 \phi$.
These observations indicate\cite{hsu,firef} that
there is a critical temperature $T_c$ at which $V_T^{res}(\phi)$
has two degenerate minima.
{}From Eqs.(\ref{appgapma})-(\ref{vresult}) it is also easy to realize that
when $e^2/\lambda>>1$ the symmetry breaking minimum $\phi_b$ is
located in the region of the $\phi$-axis  that is reliably described by
the daisy and superdaisy resummed effective
potential\footnote{\ninerm\baselineskip=11pt
As discussed in Refs.8,9,14, the daisy and superdaisy
resummed effective potential is expected to give a reliable
approximation of the full effective potential for
all $\phi \! > \! e T$.},
i.e. $\phi_b >e T_c$ (see Fig.1),
and therefore, at least
in these hypotheses,
my result indicates that the Abelian Higgs Model has a first order phase
transition.

\begin{figure}[htb]
\epsfxsize=2.8in
\centerline{\epsffile{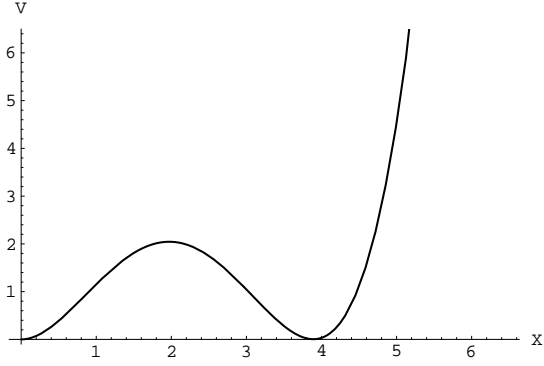}}
\caption{The Rayleigh-Ritz and high-temperature approximation
of the daisy and superdaisy resummed
effective potential at the phase transition.
In figure
$V(X) \! \equiv \! 10^{5} Re[V^{res}_T(X) \! - \! V^{res}_T(0)]/T^4$,
$X \! \equiv \! \phi / e T$,
$T \! \simeq \! T_c \! \simeq \! 8.625 m$, $e \! = \! .24$,
$\lambda \! = \! 0.01$.}
\label{fig4p7}
\end{figure}

Another interesting aspect of Eq.(\ref{vresult}) is that
the terms linear in the effective masses
have cancelled out.
In the literature there has been an extensive debate
on the possibility
that the resummation of the daisy and superdaisy diagrams
might induce contributions to the finite temperature
effective potential which are linear in the effective
masses\footnote{\ninerm\baselineskip=11pt
More precisely, it has been conjectured
that $V_T^{res} - V_{classic}- V_{one-loop}^*$,
where $V_{one-loop}^*$ is the leading one-loop
contribution (which, for example, in the case of the
Abelian Higgs Model is given by
$T^2 (2 \lambda \phi^2 /3 - 2 m^2 + 3 e^2 \phi^2)/24$),
might include terms linear in the effective masses.}.
Using the general form of the Rayleigh-Ritz approximation
with momentum independent effective masses it is easy to see
that a cancellation of linear terms always occurs.
Let me consider for simplicity a purely bosonic theory with $N$ fields,
whose tree-level masses I label $m_i$ ($i$=$1 .... N$).
For such a theory the Rayleigh-Ritz approximation
of the effective potential
(with appropriate parametrization of $G$ in terms of
effective masses $M_i$)
takes the general form
\bea
V_{R-R} &=& V_{classic}(\phi)
+{1 \over 2} \, \sumint_k ~ \ln G^{-1}(\{M \};k) \nonumber\\
& & + {1 \over 2} \, \sumint_k ~ [D^{-1}(\phi;k) G(\{M \};k) -1]
+ V_2(\phi,\{M \}) \nonumber\\
&=& V_{classic}(\phi) ~+~
\sum_i \left( {T^2 M_i^2 \over 24}
- {T M_i^3 \over 12 \pi} \, + \, ...... \right) \nonumber\\
& & + \sum_i (m_i^2 - M_i^2) \left( {T^2 \over 24}
- {T M_i \over 8 \pi} \, + \, ...... \right) ~+~ V_2(\phi,\{M \}) ~,
\label{prpotprimo}
\eea

\noindent
and the gap equations that follow from varying $V_{R-R}$ have the form
\be
M_i^2 = m_i^2 - {8 \pi \over T}
{\partial V_2(\phi,\{M \}) \over \partial M_i} \, + \, .... ~.
\label{prgap}
\ee

\noindent
Using the gap equations, $V_{R-R}$ can be written as
\bea
V_{R-R} &=& V_{classic}(\phi) ~+~
\sum_i \left( {T^2 m_i^2 \over 24}
- {T M_i^3 \over 12 \pi} \, + \, ...... \right) \nonumber\\
& & - M_i {\partial V_2(\phi,\{M \}) \over \partial M_i}
+ V_2(\phi,\{M \})
\, + \, .... ~.
\label{prpotzwei}
\eea

\noindent
Contributions linear in $M_i$ can come from the terms
$- M_i \partial V_2(\phi,\{M \}) / \partial M_i$
and $V_2(\phi,\{M \})$, but,
evidently, each linear contribution coming from $V_2(\phi,\{M \})$
is exactly cancelled by a corresponding contribution coming from
$- M_i \partial V_2(\phi,\{M \}) / \partial M_i$, leading to a combined
contribution to $V_{R-R}$ that does not include any term linear in $M_i$.
Because the derivation is independent
of the specific form of $V_2(\phi,\{M \})$,
this result
is valid to all orders (i.e. it applies to the full effective potential
and any consistent approximation of it),
and in particular it applies to the daisy and superdaisy resummed
effective potential.

The techniques discussed in this analysis of the
Abelian Higgs Model clearly apply to any gauge theory.
Using the composite operator method,
one can do better than the daisy and superdaisy resummation
by going beyond the lowest non-trivial order in the loop expansion
of $V(\phi,G)$.
Also my Rayleigh-Ritz approximation can be
improved
by using
more elaborated versions of the parameter dependent expression for $G$;
for example, one can make the
substitutions $M_{x}^2 \rightarrow  M_{x}^2 + Y_{x} {\bf k}^2$
in Eq.(\ref{ansrr})
and vary not only the $M_{x}$'s but also the additional parameters $Y_{x}$.
Numerical methods can be used both in the study of these more elaborated
versions of the Rayleigh-Ritz approximation, and
in the exact evaluation of the daisy and superdaisy
resummed effective potential
as given in Eqs.(\ref{hfahm})-(\ref{statahm}).

\vglue 0.6cm
\leftline{\twelvebf Acknowledgements}
\vglue 0.4cm
This work is
supported in part by funds provided by the U.S. Department of Energy
(D.O.E.) under contract \#DE-AC02-76ER03069.
I acknowledge support from the Istituto Nazionale di Fisica Nucleare,
Frascati, Italy.

\vglue 0.6cm
\leftline{\twelvebf References}
\vglue 0.4cm

\end{document}